\documentclass[journal=jctcce,manuscript=article]{achemso}
\setkeys{acs}{usetitle=true}

\usepackage[version=3]{mhchem} % Formula subscripts using \ce{}
\usepackage{cleveref}  %使式子多个引用函数\cref 可用
\usepackage{threeparttable}%表格下注脚
\usepackage{amsmath,amssymb}
\usepackage{graphicx,color,subfigure}
\usepackage{amsfonts}
\usepackage{comment}
\usepackage{placeins}
\usepackage{fixltx2e}

\usepackage{sectsty}
\usepackage{balance}
\usepackage{xspace}
\usepackage{algorithm,algorithmic}
\usepackage{booktabs}  %简单表格
\usepackage{makecell} %复杂表格
\usepackage{multirow} %复杂表格
\usepackage{enumitem} %罗马数字标号

\usepackage{epsfig,amssymb,amsmath,amsthm,amsfonts,amsbsy,mathrsfs}
\usepackage{graphicx,color}
\usepackage{amsmath}
\usepackage{amssymb}
\usepackage{epstopdf}
\usepackage{threeparttable}
\usepackage{graphicx} %eps figures can be used instead
\usepackage{lastpage}
\usepackage[format=plain,justification=raggedright,singlelinecheck=false,font=small,labelfont=bf,labelsep=space]{caption}
\allowdisplaybreaks

\usepackage[version=3]{mhchem} % Formula subscripts using \ce{}
\usepackage{indentfirst} % first line indent after a section
\usepackage{bm}

\newcommand{\bt}{\mathbf{\theta}}

\title[Gradient]
{Quantum Equation-of-Motion Method with Single, Double, and Triple Excitations}

\author{Yuhan Zheng}
\affiliation{Hefei National Research Center for Physical Sciences at the Microscale, University of Science and Technology of China, Hefei, Anhui 230026, China} 

\author{Jie Liu}
\email{liujie86@ustc.edu.cn} 
\affiliation{Hefei National Laboratory, University of Science and Technology of China, Hefei 230088, China}

\author{Zhenyu Li}
\email{zyli@ustc.edu.cn} 
\affiliation{Hefei National Research Center for Physical Sciences at the Microscale, University of Science and Technology of China, Hefei, Anhui 230026, China} 
\alsoaffiliation{Hefei National Laboratory, University of Science and Technology of China, Hefei 230088, China}

\author{Jinlong Yang}
\affiliation{Hefei National Research Center for Physical Sciences at the Microscale, University of Science and Technology of China, Hefei, Anhui 230026, China}
\alsoaffiliation{Hefei National Laboratory, University of Science and Technology of China, Hefei 230088, China}

\begin{document}
\begin{abstract}
The quantum equation-of-motion (qEOM) method with singles and doubles has been suggested to study electronically excited states while it fails to predict the excitation energies dominated by double excitations. In this work, we present an efficient implementation of the qEOM method with single, double and triple excitations. In order to reduce the computational complexity, we utilize the point group symmetry and perturbation theory to screen triple excitation operators, and the scaling is reduced from $N_o^6N_v^6$ to $N_o^5N_v^5$. Furthermore, we introduce a perturbation correction to the excitation energy to account for the effect of ignored triple excitation operators. We apply this method to study challenging cases, for which the qEOM-SD method exhibits large errors, such as the 2\,$^1\Delta$ excited state of $\rm{CH}^+$ and the 2\,$^1\Sigma$ state of $\rm{H}_8$ molecule. Our new method yields the energy errors less than 0.18 eV.

\end{abstract}

\section{Introduction} \label{sec:Introduction}

Accurate prediction of excited-state properties has broad implications for many important real-world scenarios, such as the design of organic light-emitting diodes and the study of photosynthesis reaction mechanisms, while posing a significant challenge for classical computing with (near-) exact approaches due to unfavorable computational scaling~\cite{Pul11,SzaMulGid12,BaiKelRei22,SchEva17}. The emergence of quantum computing provides a new paradigm to solve the Schr\"odinger equation in an accurate and efficient manner by utilizing quantum superposition and entanglement, where $N$ qubits can represent a quantum state of $2^N$ dimensions~\cite{aspuru2005simulated,peruzzo2014variational,mcardle2020quantum,su2021fault,o2016scalable}. It is therefore necessary to explore carrying out excited-state simulations on quantum computers.

In recent years, many quantum algorithms have emerged for finding electronically excited states of a given Hamiltonian~\cite{parrish2019quantum, nakanishi2019subspace, ryabinkin2018constrained,   stair2020multireference, jones2019variational, tilly2020computation}.
In this regard, the Hamiltonian simulation algorithms, including quantum phase estimation~\cite{BauLiuByl21,KanBauKri22}, quantum Krylov subspace expansion~\cite{stair2020multireference} and quantum Lanczos recursion~\cite{baker2021lanczos}, are a straightforward way to implement excited-state simulations. However, these algorithms require to apply the Hamiltonian $\hat{H}$ or $e^{-i \hat{H} t}$ many times onto a quantum state, resulting in deep quantum circuits. Additionally, these algorithms are generally sensitive to noise in quantum systems so that it is difficult to obtain reliable results due to error accumulation. As such, they are considered as long-term algorithms to implement on a fault-tolerant quantum computer. 

Given near-term quantum devices limited by the coherence time and the fidelity of quantum gates~\cite{preskill2018quantum,2020Noise}, variational quantum eigensolver (VQE)~\cite{peruzzo2014variational}, known as a hybrid quantum-classical algorithm that has features of low circuit depth and robust noise resilience, is one of the most widely used algorithms for electronic structure simulations. Various excited-state methods have been also developed based on the VQE algorithm. One simple way to extend the VQE to find excited states is to optimize parameterized quantum circuits (PQC) under certain constrained conditions. For examples, the variational quantum deflation algorithm~\cite{chan2021molecular,higgott2019variational} adds a penalty function to the original Hamiltonian in the VQE so that the $(k+1)$th eigenstate is optimized to be orthogonal to all previously computed $k$ eigenstates. The folded spectrum method~\cite{peruzzo2014variational} minimizes an objective function $\langle (\hat{H}-E_\lambda)^2 \rangle$ to find a certain eigenstate with the energy closest to $E_\lambda$. In contrast to the above two algorithms, the subspace search VQE algorithm \cite{nakanishi2019subspace} tries to obtain the lowest-lying $k$ eigenstates simultaneously, which are represented by applying an ansatz circuit onto $k$ orthogonal initial states. Witness-assisted variational eigenspectra solver~\cite{santagati2018witnessing} utilizes an additional qubit, known as a witness qubit, to monitor the quantum circuit's output and determines whether it is the target excited state. The major challenge faced by PQC-based approaches is the high-dimension nonlinear optimization problem as the number of circuit parameters increases. In case of excited-state calculations, it is also a challenge task to find the target state because the optimization procedure may get stuck in other states~\cite{lee2018generalized,nakanishi2019subspace}.

Alternatively, one can employ diagonalization methods to extract multiple low-lying eigenstates of a given Hamiltonian. A typical example is the quantum subspace expansion (QSE) algorithm, which prepares an approximate ground state in a small active space and then expands the ground state and excited states with the same subspace~\cite{mcclean2017hybrid}. In contrast, Ollitrault et al.~\cite{ollitrault2020quantum} proposed the quantum equation-of-motion (qEOM) method for excited-state calculations. The qEOM employs the unitary coupled-cluster (UCC) ansatz to represent the ground state and expands the excited-state wave functions as a linear combination of basis functions commonly generated by applying single and double excitation operators onto the ground state. Since the qEOM fails to satisfy the killer condition, it leads to significant errors when used to calculate vertical ionization potentials (IPs) and electron affinities (EAs). To address this issue, qEOM was extended to quantum self-consistent equation-of-motion (q-sc-EOM)~\cite{asthana2023quantum}. In the q-sc-EOM, the excitation operator manifold is rotated, meaning that the excitation operators undergo a similarity transformation using the optimized UCC operator to satisfy the killer condition. In addition, the q-sc-EOM employs the adaptive derivative-assembled pseudotrotter (ADAPT) ansatz to generate the accurate ground-state wave function for small molecules. %In addition, Fan et al. \cite{fan2021equation} combined the EOM theory and the adaptive derivative-assembled pseudotrotter (ADAPT) VQE algorithm to simulate band structures of periodic systems, where a projected excitation operator formalism was introduced to satisfy the killer condition. 
  
Analogous to the classical EOM approaches, the computational accuracy of quantum EOM approaches also depends on the truncation of excitation operators. As stated in previous works, the EOM coupled cluster with singles and doubles (CCSD) method introduced significant errors when computing excited states dominated by double excitation components and excited-state potential energy surfaces (PESs) involving bond breaking. For example, the energy error for the $2\,^1 \Delta$ state of the $\rm CH^+$ ion is as high as 3.946 eV, while the energy error for the $1\,^1 \Delta_g$ state of the $\rm C_2$ molecule reaches 2.068 eV ~\cite{piecuch2004method,wloch2005extension}. In this work, we found that the errors of the qEOM unitary CCSD (qEOM-UCCSD) are within the same order of magnitude as EOM-CCSD. Thus, in order to improve the computational accuracy, it is necessary to consider higher-order excitation operators. 

In this work, we extend the qEOM-SD algorithm to include triple excitations via many-body perturbation theory, resulting in the qEOM-SD(t) method. This method leverages orbital symmetries and perturbation theory to identify and eliminate redundant excitation operators. Furthermore, it replaces the effects of discarded excitation operators on excitation energies with their associated first-order perturbation energies. We test this method on $\rm CH^+$, HF, and $\rm H_8$, demonstrating significant improvements over the qEOM-SD method. Additionally, we extend the QSE algorithm to include triple excitations, introducing the QSE-SD(t) method, and compare it with the qEOM-SD(t).

\section{Methodology} \label{sec:Methodology}
\subsection{ADAPT-VQE algorithm}
The ADAPT-VQE algorithm is utilized to generate the ground-state wave function~\cite{grimsley2019adaptive}
\begin{equation}\label{eq:wf}
    |\psi_{0}\rangle = |\psi_{\rm VQE}\rangle = \hat{U}(\bt) |\psi_{\rm HF}\rangle =  \prod \limits_{k=0}^K \hat{U}(\theta_k) | \psi_{\rm HF}\rangle
\end{equation}
where $|\psi_{\rm HF}\rangle$ is the Hartree-Fock state and the unitary transformation $\hat{U}(\theta_k)$ is in the form of  
\begin{equation}
    \hat{U}(\theta_k) = e^{\theta_k \hat{\tau}_k}
\end{equation}
with $\hat{\tau}_k = \hat{T}_k - \hat{T}_k^\dagger$ being an anti-Hermitian operator. The operator pool $\mathcal{O} = \{\hat{T}_k \}$ is often assumed to be composed of all single and double excitation operators
\begin{gather}
\begin{split}
&\hat{T}_i^a=\hat{a}_{a}^{\dagger} \hat{a}_{i}\\
&\hat{T}_{ij}^{ab} = \hat{a}_{a}^{\dagger}\hat{a}_{b}^{\dagger} \hat{a}_{j} \hat{a}_{i}
\end{split}
\end{gather}
In this work, we employ single and double excitation operators in the ADAPT-VQE calculations. The wave function of Eq.~\eqref{eq:wf} is iteratively determined 
\begin{gather}
|\psi^{(k)}\rangle = e^{\theta_k \hat{\tau}_k }|\psi^{(k-1)}\rangle,
\end{gather}
where the operator $ \hat{\tau}_k$ has the largest residual gradient $g_k = \max \limits_{\mu \in \mathcal{O}} g_\mu$ with
\begin{equation}
    g_\mu = \left. \frac{\partial \langle \psi^{(k-1)} |e^{\theta \hat{\tau}^\dagger } \hat{H} e^{\theta \hat{\tau} }| \psi^{(k-1)} \rangle}{\partial \theta} \right|_{\hat{\tau}=\hat{\tau}_\mu,\theta=0}
\end{equation}
with respect to the wave function $|\psi^{(k-1)}\rangle$. In each iteration, variational parameters are optimized using the variational principle
\begin{gather}
E^{(k)} = \mathop{\min}_{\bm{\theta}} E^{(k)}(\bm{\theta}) = \frac{\langle \psi^{(k)}(\bm{\theta})|H|\psi^{(k)}(\bm{\theta})\rangle}{\langle \psi^{(k)}(\bm{\theta})|\psi^{(k)}(\bm{\theta})\rangle} 
\end{gather} 
When the norm of residual gradients is smaller than a threshold $\epsilon$
\begin{equation}
    \sqrt{\sum_\mu |g_\mu|^2} < \epsilon,
\end{equation}
the iterative optimization procedure is finished.

\subsection{Quantum equation-of-motion method}
The EOM theory introduce an excitation operator $\hat{R}_m$ to define the $m$th excited state wave function
\begin{equation}\label{eq1}
|\psi_m\rangle = \hat{R}_m|\psi_{0}\rangle,
\end{equation}
and the excitation operator satisfies the EOM equation
\begin{equation}\label{eq:eom}
[\hat{H},\hat{R}_m] = \omega_{m}\hat{R}_m
\end{equation}
where $\omega_{m}$ is the $\textit{m}$th vertical excitation energy. The excitation operator $\hat{R}_m$ should in principle satisfy the killer condition or the vacuum annihilation condition (VAC)
\begin{equation}\label{eq:VAC}
    \hat{R}_m^\dagger |\psi_{\rm 0}\rangle = 0,
\end{equation}
implying that the ground state cannot be de-excited. In the EOM-CC theory, an effective Hamiltonian
\begin{equation}
    \bar{H} = e^{-\hat{T}} \hat{H} e^{\hat{T}}
\end{equation}
is defined and the reference state $|\psi_0\rangle =|\psi_{\rm HF}\rangle$. Here, $\hat{T} = \sum_{ai} \hat{T}^a_i + \frac{1}{4} \sum_{abij} \hat{T}^{ab}_{ij}+\cdots$. As such, the VAC is always satisfied if the excitation operators are restricted to promote electrons from occupied to virtual orbitals. 

Inspired by the EOM-CC theory, one can construct an effective Hamiltonian using the unitary transformation $\hat{U}$ in Eq.~\eqref{eq:wf} as
\begin{equation}\label{eq:Heff}
    \bar{H} =\hat{U}^{\dagger} \hat{H} \hat{U}
\end{equation}
and the ground-state energy is rewritten as
\begin{equation}
    E_0 = \langle \psi_{\rm HF} | \bar{H} | \psi_{\rm HF} \rangle.
\end{equation}
As such, the $\textit{m}$th excited state is defined as 
\begin{equation}
    |\psi_m\rangle = \hat{R}_m |\psi_{\rm HF}\rangle
\end{equation}
with respect to the effective Hamiltonian of Eq.~\eqref{eq:Heff}. In this work, $ \hat{R}_m$ is approximated as the linear combination of single, double and triple excitations
\begin{equation}\label{eq:R}
    \begin{split}
        \hat{R}_m &= \sum_{ia}\gamma_i^a\hat{a}_a^{\dag}\hat{a}_i + \sum_{ijab}\gamma_{ij}^{ab}\hat{a}_a^{\dag}\hat{a}_b^{\dag}\hat{a}_i\hat{a}_j +
\sum_{ijkabc}\gamma_{ijk}^{abc}\hat{a}_a^{\dag}\hat{a}_b^{\dag}\hat{a}_c^{\dag}\hat{a}_i\hat{a}_j\hat{a}_k\\ &= \sum_I \gamma_I\hat r_I
    \end{split}
\end{equation}
where $i$, $j$ and $k$ represent occupied spin orbitals, $a$, $b$ and $c$ represent virtual spin orbitals. We name this method qEOM-SDT. It is clear that the VAC is also satisfied when one consider the Hartree-Fock state as the reference state.

Analogous to the EOM-CC theory, the working equation of the qEOM theory is formulated as
\begin{equation}\label{eq3}
\langle \psi_{\rm HF} |\hat{r}_I^{\dag}[\bar{H},\hat{R}_m]|\psi_{\rm HF} \rangle = \omega_{m} \langle \psi_{\rm HF}| \hat{r}_I^{\dag} \hat{R}_m| \psi_{\rm HF} \rangle
\end{equation}
After some rearrangements, the eigenvalue equation is formulated as
\begin{equation}\label{eq:eom-ccsdt}
\boldsymbol{M\Gamma}_m=\omega_{m}\boldsymbol{\Gamma}_m
\end{equation}
with $\omega_{m}$ is the vertical excitation energy defined as the energy difference between the ground state and the $\textit{m}$th excited state.
The matrix elements of $\boldsymbol M$ is
\begin{equation}\label{eq:M}
\boldsymbol{M}_{I,J}=\langle {\psi_{\rm HF}}|\hat{r}_I^\dag  (\bar{H}-E_0) \hat{r}_J |{\psi_{\rm HF}} \rangle
\end{equation}
The coefficients $\boldsymbol{\Gamma}_m=\{\gamma_i^a,\gamma_{ij}^{ab},\gamma_{ijk }^{abc}\}$ in $\hat{R}_m$ can be obtained by diagonalizing Eq.~\eqref{eq:eom-ccsdt}. Note that Eq.~\eqref{eq:eom-ccsdt} is the same as the eigenvalue equation proposed in the q-sc-EOM approach~\cite{asthana2023quantum}.

\subsection{qEOM-SDt method}\label{section:Reducing operators by symmetry}
%If we consider all operators in the set $\mathbf{U}$, the resulting quantum circuit used to calculate elements $\boldsymbol{M}_{I,J}$ would become deep. Similarly, if we consider all operators in the set $\mathbf{R}$, a large number of elements would need to be measured. Both of these pose computational challenges. Therefore, in the following, we introduce a method, named qEOMCSD+t, which employ geometric symmetry and perturbation theory to mitigate the computational complexity associated with these two sets of operators.

%The EOM method consists of three components: symmetry-based reduction of operators in $\mathbf{U}$, symmetry-based reduction of operators in $\mathbf{R}$, and perturbation theory for ranking high-order operators in $\mathbf{R}$.
The number of triple excitation operators scales as $N_o^3N_v^3$ with $N_o$ and $N_v$ being the number of occupied and virtual spin orbitals. As such, the major computational bottleneck of qEOM-SDT results from evaluating the matrix elements of $\boldsymbol M$ related to the triple excitations, scaling as $N_o^6N_v^6$. Here, we exploit molecular orbital symmetry and the perturbation theory to reduce the computational complexity.

\noindent\textbf{(1) Symmetry Preserving in ADAPT-VQE}

%We utilize the point group symmetry to reduce operators within the ADAPT ansatz. 
As discussed in Ref. \cite{cao2022progress}, when the ground state is non-degenerate, its irreducible representation (Irrep) is in principle the same as the reference state. As such, in the ADAPT-VQE calculation, we have
\begin{equation}\label{eq6}
{\rm Irrep}(|{\psi_{\rm VQE}}\rangle)={\rm Irrep}(|{\psi_{\rm HF}}\rangle)
\end{equation}
Since the Hartree-Fock state cannot be de-excited, the above condition is equivalent to the following one:
\begin{equation}\label{eq9}
{\rm Irrep}(\hat{T}_i|{\psi_{\rm HF}\rangle})=\rm Irrep(|\psi_{HF}\rangle)
\end{equation}
This condition can serve as a guideline for building the operator pool in the ADAPT-VQE algorithm. The detailed derivation of Eq.~\eqref{eq9} is provided in Appendix.

\noindent\textbf{(2) Block Diagonalization}

The $\boldsymbol{M}$ is the matrix representation of $\bar{H}$ in the subspace $\{\hat{r}_1 |{\psi_{\rm HF}} \rangle,\hat{r}_2 |{\psi_{\rm HF}} \rangle,\hat{r}_3 |{\psi_{\rm HF}} \rangle, \cdots\}$. Here, $\hat{r}_I \in \hat{\mathbf{R}}$, where $\hat{\mathbf{R}} = \hat{\mathbf{R}}_1 \cup \hat{\mathbf{R}}_2 \cup \hat{\mathbf{R}}_3$,  $\hat{\mathbf{R}}_1 = \{\hat{a}_a^\dagger \hat{a}_i\}$, $\hat{\mathbf{R}}_2 = \{\hat{a}_a^\dagger  \hat{a}_b^\dagger \hat{a}_i \hat{a}_j\}$ and $\hat{\mathbf{R}}_3 = \{\hat{a}_a^\dagger \hat{a}_b^\dagger \hat{a}_c^\dagger \hat{a}_i \hat{a}_j \hat{a}_k\}$.
As the unitary transformation $\hat{U}$ determined in the VQE will not change the irreducible representation of the reference configuration state, it belongs to the totally symmetric irreducible representation, such as $A_\mathrm{g}$ in $D_{2h}$ group and $A_1$ in $C_{2v}$ group. According to the group theory~\cite{cotton1991chemical}, the Hamiltonian $\hat{H}$ belongs to the totally symmetric irreducible representation. If bra $\hat{U} \hat{r}_I |{\psi_{\rm HF}} \rangle$ and ket $\hat{U} \hat{r}_J |{\psi_{\rm HF}} \rangle$ on both sides of $\hat{H}$ have different irreducible representations, the matrix element $\boldsymbol{M}_{I,J}$ is zero. This reveals that we can implement a block diagonalization of the eigenvalue equation of Eq.~\eqref{eq:eom-ccsdt} for excited states per irreducible representation 
\begin{equation}\label{eq20}
{\rm Irrep}(\hat{r}_I|{\psi_{\rm HF}\rangle}) = {\rm Irrep}(|\psi_m\rangle)  
\end{equation}
\\
\noindent\textbf{(3) Screening Triple Excitation Operators via the Perturbation Theory}\label{sec:screening}

We assume that the zero-order of $m$th excited state for the effective Hamiltonian $\bar{H}$ is only formed by single and double excitations
\begin{equation}\label{eq21}
{|\psi_{m}^{(0)}\rangle}= \sum_{\substack{\hat{r}_I\in {\hat{\mathbf{R}}_{1}\cup\hat{\mathbf{R}}_{2}}}} \gamma_I\hat{r}_I|\psi_{\rm HF}\rangle
\end{equation}
and the first-order wave function is defined as
\begin{equation}\label{eq22}
{|\psi_{m}^{(1)}\rangle}= \sum_{\substack{\hat{r}_I\in {\hat{\mathbf{R}}_{3}}}} \gamma_I\hat{r}_I|\psi_{\rm HF}\rangle
\end{equation}
So
\begin{equation}\label{eq23}
{|\psi_{m}\rangle}=|\psi_{m}^{(0)}\rangle+|\psi_{m}^{(1)}\rangle
\end{equation}
\begin{equation}\label{eq232}
E_m \approx E_m^{(0)} + E_m^{(1)}
\end{equation}
And $E_m^{(0)} = \langle {\psi_{m}^{(0)}}| \bar{H}|\psi_{m}^{(0)} \rangle$. Substituting Eq.~\eqref{eq23} and \eqref{eq232} into $\bar{H}|\psi_{m}\rangle=E_m|\psi_{m}\rangle$, and ignoring high-order term $E_m^{(1)}|\psi_m^{(1)}\rangle$, we get
\begin{equation}\label{eq24}
\bar{H}(|\psi_{m}^{(0)}\rangle+|\psi_{m}^{(1)}\rangle)=(E_m^{(0)} + E_m^{(1)})|\psi_{m}^{(0)}\rangle+E_m^{(0)}|\psi_{m}^{(1)}\rangle
\end{equation}
Left-multiplying Eq. \eqref{eq24} by $\langle \psi_{\rm HF}|\hat{r}_J^{\dagger}$, where $\hat{r}_J \in \hat{\mathbf{R}}_{3}$ is a crucial point to note, we obtain
\begin{equation}\label{eq25}
\langle \psi_{\rm HF}|\hat{r}_J^{\dagger}\bar{H}|\psi_{m}^{(0)}\rangle+\langle \psi_{\rm HF}|\hat{r}_J^{\dagger}\bar{H}(\sum_{\substack{\hat{r}_I\in {\hat{\mathbf{R}}_{3}}}} \gamma_I\hat{r}_I)|\psi_{\rm HF}\rangle=E_m^{(0)} \gamma_J
\end{equation}
since $\langle \psi_{\rm HF}|\hat{r}_J^{\dagger}|\psi_m^{(0)}\rangle=0$. Considering the diagonal
approximation, i.e., $\langle \psi_{\rm HF}|\hat{r}_J^{\dagger}\bar{H}\hat{r}_I|\psi_{\rm HF}\rangle \approx \delta_{IJ}\langle \psi_{\rm HF}|r_J^{\dagger}\bar{H}\hat{r}_J|\psi_{\rm HF}\rangle$, 
\begin{equation}\label{eq26}
\gamma_J = \frac{\langle \psi_{\rm HF}|\hat{r}_J^{\dagger}\bar{H}|\psi_{m}^{(0)}\rangle}{E_m^{(0)}-\langle \psi_{\rm HF}|\hat{r}_J^{\dagger}\bar{H}\hat{r}_J|\psi_{\rm HF}\rangle} 
\end{equation}
Left-multiplying Eq. \eqref{eq24} by $\langle \psi_{m}^{(0)}|$, we obtain
\begin{equation}\label{eq25}
\langle \psi_{m}^{(0)}|\bar{H}|\psi_{m}^{(0)}\rangle+\langle \psi_{m}^{(0)}|\bar{H}(\sum_{\substack{\hat{r}_I\in {\hat{\mathbf{R}}_{3}}}} \gamma_I\hat{r}_I)|\psi_{\rm HF}\rangle=E_m^{(0)}+E_m^{(1)}
\end{equation}
Therefore, 
\begin{equation}\label{eq25}
\begin{split}
E_m^{(1)}&=\langle \psi_{m}^{(0)}|\bar{H}(\sum_{\substack{\hat{r}_I\in {\hat{\mathbf{R}}_{3}}}} \gamma_I\hat{r}_I)|\psi_{\rm HF}\rangle\\
&=\sum_{\substack{\hat{r}_I\in {\hat{\mathbf{R}}_{3}}}} \frac{|\langle \psi_{\rm HF}|\hat{r}_I^{\dagger}\bar{H}|\psi_{m}^{(0)}\rangle|^{2}}{E_m^{(0)}-\langle \psi_{\rm HF}|r_I^{\dagger}\bar{H}\hat{r}_I|\psi_{\rm HF}\rangle}
\end{split}
\end{equation}
Thus, one can define weight coefficients
\begin{equation}\label{eq:W}
W_I=\frac{|\langle \psi_{\rm HF}|r_I^{\dagger}\bar{H}|\psi_{m}^{(0)}\rangle|^{2}}{E_m^{(0)}-\langle \psi_{\rm HF}|\hat{r}_I^{\dagger}\bar{H}\hat{r}_I|\psi_{\rm HF}\rangle}
\end{equation}
to determine the importance of triple excitation operators. Given a threshold $\epsilon_t$, only triple excitation operators $\hat{r}_I$ with $|W_I|>\epsilon_t$ are involved in the qEOM-SDt calculations. 

\noindent\textbf{(4) Perturbation Correction}\label{Target Energy}

After step (2) and (3), the dimension of the $\boldsymbol{M}$ matrix is significantly reduced. We denote the reduced matrix as $\boldsymbol{M}'$. An approximate excitation energy is obtained by diagonalizing the eigenvalue equation
\begin{equation}\label{eq:eom-ccsdt'}
\boldsymbol{M'\Gamma}_m'=\omega_{m}'\boldsymbol{\Gamma}_m'
\end{equation}

The influence of the ignored triple excitation operators on the excitation energy can be considered using the perturbation theory
\begin{equation}\label{eq:eom-ccsdt'}
w = w' + \sum_{\substack{\hat{r}_I\in {\hat{\mathbf{R}}_{3}\backslash\hat{\mathbf{R}}_{3}'}}} \frac{|\langle \psi_{\rm HF}|\hat{r}_I^{\dagger}\bar{H}|\psi_{m}^{(0)}\rangle|^{2}}{E_m^{(0)}-\langle \psi_{\rm HF}|r_I^{\dagger}\bar{H}\hat{r}_I|\psi_{\rm HF}\rangle}
\end{equation}

\begin{figure}[H] 
    \centering
    \makebox[\textwidth][c]{\includegraphics[width=0.38\textwidth]{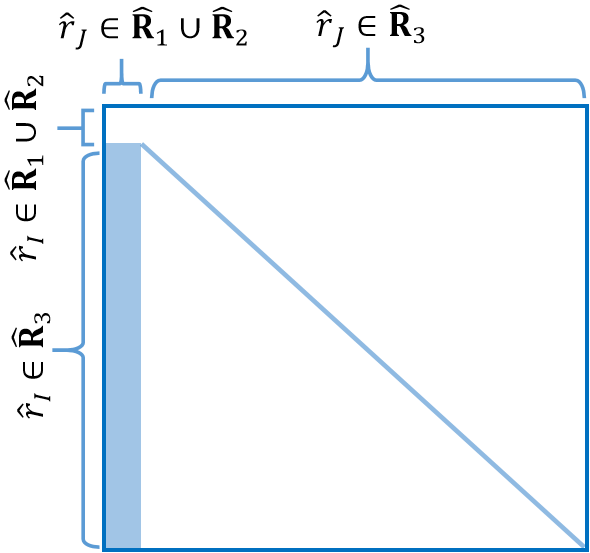}}
    \caption{\centering{Matrix elements of $\boldsymbol{M}$ matrix required to be measured during the operator selection process are colored in blue.
}}
    \label{fig8}
\end{figure}

 In the original qEOM-SDT method, the computational complexity for measuring the $\boldsymbol{M}$ matrix elements scales as $N_o^6N_v^6$. In the qEOM-SDt method, the computational bottleneck results from selecting triple excitation operators. In the screening procedure, the matrix elements $\langle \psi_{\rm HF}|\hat{r}_I^{\dagger}\bar{H}|\psi_{m}^{(0)}\rangle$ and $\langle \psi_{\rm HF}|\hat{r}_I^{\dagger}\bar{H}\hat{r}_I|\psi_{\rm HF}\rangle$ with ${\hat{r}_I\in {\hat{\mathbf{R}}_{3}}}$ are calculated on a quantum computer. According to Eq. \eqref{eq21}, $|\psi_{m}^{(0)}\rangle$ is in the form of a linear combination of $\hat{r}_J|\psi_{\rm HF}\rangle$, where $\hat{r}_J\in {\hat{\mathbf{R}}_{1}\cup\hat{\mathbf{R}}_{2}}$. Consequently, only the non-diagonal elements $\{\langle \psi_{\rm HF}|\hat{r}_I^{\dagger}\bar{H}\hat{r}_J|\psi_{\rm HF}\rangle|\hat{r}_I\in {\hat{\mathbf{R}}_{3}}, \hat{r}_J\in {\hat{\mathbf{R}}_{1}\cup\hat{\mathbf{R}}_{2}}\}$ and the diagonal elements of the $\boldsymbol{M}$ matrix are calculated, as indicated by the blue area in Fig. \ref{fig8}. The number of double and triple excitation operators scales as $N_o^2N_v^2$ and $N_o^3N_v^3$, respectively. The computational scaling is thus reduced to $N_o^5N_v^5$, instead of $N_o^6N_v^6$ in the standard qEOM-SDT. The strategy for measuring the non-diagonal elements can be found in Ref. \cite{asthana2023quantum}.

%\subsection{Computational cost estimation}\label{Computer cost estimation}
%The symmetry reduction procedure of $\mathbf{R}$ can be simple conducted on a classical computer. Here, we focus on the computational cost of calculating the weight coefficients of triple excitation operators in Eq.~\eqref{eq:W}. This task requires measuring matrix elements of the $\boldsymbol M$ matrix on a quantum computer. These elements include those associated with single and double excitations, diagonal elements, and those associated with both triple excitations and single or double excitations. Since the $\boldsymbol M$ matrix is Hermitian, it is only necessary to compute the mentioned elements in the lower triangle, as depicted by the blue area in Fig. \ref{fig8}.
%\noindent The computational scaling is significantly reduced to $N_o^5N_v^5$, compared to the $N_o^6N_v^6$ scaling required for measuring all matrix elements of the $\boldsymbol{M}$ matrix when applying the qEOM-SDT method.

%After symmetry and perturbation reduction, the total number of single, double, and triple excitations is greatly reduced. This results in shorter quantum circuits for measuring the elements of the $\boldsymbol{M}$ matrix, and a significant decrease in the number of elements as well.

\section{Results and discussion} \label{sec:Results}
All the calculations are performed using the high-performance quantum emulating software Q$^2$Chemistry~\cite{JUSTC-2022-0118}. The molecule orbitals and one- and two-electron integrals are evaluated using the PySCF~\cite{sun2018pyscf}. The fermion-to-qubit mapping is executed through the Jordan-Wigner transformation using the OpenFermion~\cite{mcclean2020openfermion}. All optimizations in the VQE method utilize the Broyden-Fletcher-Goldfarb-Shanno (BFGS) algorithm implemented in the SciPy~\cite{virtanen2020scipy}, with a convergence criteria of $\epsilon=10^{-3}$ for the norm of residual gradients. In the following, we indicate the qEOM-SDT methods before and after operator screening as qEOM-SDT and qEOM-SDt, respectively. A perturbation correction, as discussed in Section \ref{section:Reducing operators by symmetry} (4), is further introduced to the qEOM-SDt, resulting in the method called qEOM-SD(t).

\subsection{2\,$^1 \Delta$ excited state of $\rm CH^+$}\label{Symmetry reducing EOM for small molecules}
The EOM-CCSD method performs poorly in predicting the second $^1 \Delta$ excitation energy of $\rm CH^+$, with an energy error exceeding 3.5 eV. It is important to include the triple excitation operators in the EOM method to accurately describe this state~\cite{kowalski2004new}. The energy of the 2\,$^1 \Delta$ excited state is calculated at twice the equilibrium bond length ($2R_e = 2\times 2.137 13\enspace\rm{bohr}$) using the 6-31g basis set. $\rm CH^+$ possesses $C_{\infty v}$ symmetry, but in the calculations, a lower $C_{2v}$ symmetry is used. 

\begin{figure}[H] 
    \centering
    \makebox[\textwidth][c]{\includegraphics[width=0.7\textwidth]{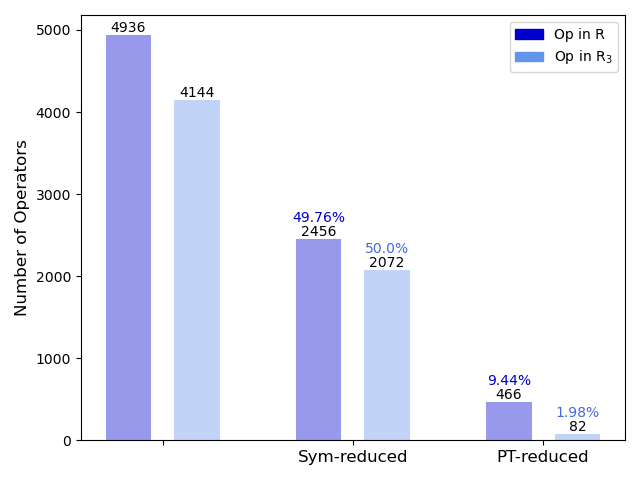}}
    \caption{\centering{Reduction of the number of operators in the operator pool $\hat{\mathbf{R}}$ and $\hat{\mathbf{R}}_3$ for the qEOM-SD(t) method when $\epsilon_t = 2.2 \times 10^{-5}$.
}}
    \label{zhuzi}
\end{figure}

The excitation energy deviations of qEOM-SDT, qEOM-SDt, qEOM-SD(t) and qEOM-SD with respect to the full configuration interaction (FCI) method are provided in Table \ref{table4}. Analogous to the EOM-CCSD, for predicting the excitation energy of 2\,$^1 \Delta$, the qEOM-SD notably overestimates it by 3.90 eV. However, incorporating 82 (1.98\%) triple excitation operators in qEOM-SDt reduces the errors by one orders of magnitude. As the number of retained operators increases, the errors of qEOM-SDt and qEOM-SD(t) approach that of qEOM-SDT, which is 0.11 eV. However, the rate of improvement in accuracy gradually slows down. The accuracy of qEOM-SD(t) consistently outperforms that of qEOM-SDt, but the difference between the two gradually diminishes. When $\epsilon_t = 2.2 \times 10^{-5}$, the number of operators in the $\hat{\mathbf{R}}=\hat{\mathbf{R}}_1 \cup \hat{\mathbf{R}}_2 \cup \hat{\mathbf{R}}_3$ and $\hat{\mathbf{R}}_3$, both before and after the operator screening are shown in Fig. \ref{zhuzi}. It demonstrates that the reduction in the number of operators based on symmetry exceeds 40\% for $\hat{\mathbf{R}}$ and $\hat{\mathbf{R}}_3$, and similarly for perturbation, indicating significant effects for both approaches.
%the energy error calculated by the qEOM-SDt method is within the same order of magnitude as the qEOM-SDT method and surpasses the qEOM-SD method by one orders of magnitude. Furthermore, by including 0.54\% of the triple excitation operators, we can cover more than 90\% of the energy correction contributed by all triple excitations. 

\begin{table}[!htbp]
\caption{\centering Energy deviations of the $2 ^1\Delta$ excited state computed with qEOM-SDt,qEOM-SD(t), qEOM-SDT and qEOM-SD methods for the CH$^+$ molecule at twice the equilibrium bond length. The reference values are computed with the FCI method.}
\label{table4}
\centering
\resizebox{\linewidth}{!}{
    \begin{tabular}{ccccccc}
       \toprule
       \midrule
         $\epsilon_t$& Op in $\hat{\mathbf{R}}_3$ after (before) { } & $\%$ & $\Delta E_{\rm qEOM-SDt}$&$\Delta E_{\rm qEOM-SD(t)}$ &$\Delta E_{\rm qEOM-SDT}$&$\Delta E_{\rm qEOM-SD}$\\
       \midrule
        $2.2 \times 10^{-5}$& 82 (4144) & $1.98\%$ & 0.2004& 0.1736& \multirow{3}{*}{0.1075} &\multirow{3}{*}{3.9194}\\
        $3.0 \times 10^{-6}$& 182 (4144) & $4.39\%$ & 0.1278& 0.1243& {} &{}\\
        $2.55 \times 10^{-7}$& 282 (4144) & $6.81\%$ & 0.1196& 0.1191& {} &{}\\
       \bottomrule  
    \end{tabular}}
\begin{tablenotes} %添加此处
    \item The unit of energy difference is eV. 
 \end{tablenotes} %添加此处
\end{table}

\subsection{2\,$^1\Sigma$ excited state of HF}\label{Potential energy surface of HF}
An accurate description of the bond breaking for the HF molecule is a challenge task for traditional electronic structure methods. Here, we apply the qEOM methods to study the second $^1\Sigma$ state of the HF molecule at different bond lengths with the 6-31G basis, freezing the lowest molecular orbital. Although HF belongs to the $C_{\infty v}$ symmetry group, we use the lower $C_{2v}$ symmetry for convenience. In this symmetry, both the HF state and the 2\,$^1\Sigma$ excited state belong to the $A_1$ irrep. 

\begin{figure}[H] 
    \centering
    \makebox[\textwidth][c]{\includegraphics[width=1.0\textwidth]{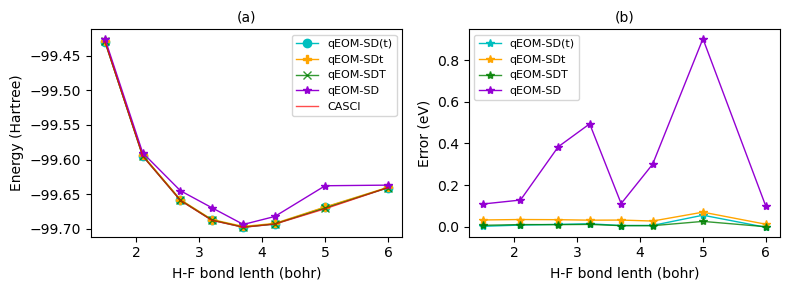}}
    \caption{\centering{(a) Potential energy surfaces (b) energy error as a function of the H-F bond length computed using CASCI, qEOM-SD, qEOM-SDt, qEOM-SD(t) and qEOM-SDT methods.
}}
    \label{fig7}
\end{figure}

Fig. \ref{fig7} shows the energy differences as a function of the H-F bond length $R_{\rm H-F}$ between the qEOM-SDT, qEOM-SDt, qEOM-SD(t) and qEOM-SD methods compared to complete active space configuration interaction (CASCI). The results indicate that the qEOM-SDt and qEOM-SD(t) methods are sufficiently accurate compared to the exact diagonalization method CASCI. The qEOM-SD method fails to accurately capture the electron correlation of the 2\,$ ^1\Sigma$ state at the large bond length. Specifically, when $R_{\rm H-F}=5$ bohr,  the error exceeds 0.9 eV. In contrast, the qEOM-SDt and qEOM-SD(t) methods exhibit much smaller errors of approximately 0.07 eV and 0.06 eV, respectively.

Table \ref{table2} illustrates that the qEOM-SDt and qEOM-SD(t) methods, using a maximum of only 61 (1.36\%) triple excitation operators, accurately recover the correlation energy in the 2\,$^1\Sigma$ excited state, with average errors of 0.03 eV and 0.01 eV, respectively. Additionally, when $R_{\rm H-F}=1.5$ and $2.1$ bohr the qEOM-SD(t) results outperform qEOM-SDT. However, as the threshold $\epsilon_t$ increases, the precision of qEOM-SD(t) and EOM-SDt will gradually approach that of qEOM-SDT. While other methods tend to overestimate the excitation energy, qEOM-SD(t) underestimates it when $R_{\rm H-F}=6.0$ bohr. Generally, qEOM-SD(t) yields better results than qEOM-SDt. Note that the excitation state accuracy obtained from qEOM-SDt and qEOM-SD(t) calculations based on VQE may surpass the ground state accuracy obtained from VQE calculations. Similar findings are also mentioned in Ref. \cite{ollitrault2020quantum} for qEOM-SD.

\begin{table}[!htbp]
    \caption{\centering Energy deviations of the $2 ^1\Sigma$ excited state computed with qEOM-SDt, qEOM-SD(t), qEOM-SDT and qEOM-SD methods for the HF molecule at different bond lengths. The reference values are computed with the CASCI method.}
    \label{table2}
    \centering
    \resizebox{\linewidth}{!}{
    \begin{tabular}{cccccccc}
       \toprule
       \midrule
         $R_{\rm H-F}$& \textbf{Op} in $\hat{\mathbf{R}}_3$ after(before)  &$\%$ & $\Delta E_{\rm qEOM-SDt}$ & $\Delta E_{\rm qEOM-SD(t)}$ &$\Delta E_{\rm qEOM-SDT}$&$\Delta E_{\rm qEOM-SD}$\\
       \midrule
       1.5 & 40 (4480) &0.89\%& 0.0324 & 0.0019 & 0.0067 & 0.1089\\
       2.1 & 48 (4480) &1.07\%& 0.0345 & 0.0079 & 0.0103 & 0.1280 \\
       2.7 & 61 (4480) &1.36\%& 0.0337 & 0.0110 & 0.0100 & 0.3837 \\
       3.2 & 43 (4480) &0.96\%& 0.0314 & 0.0141 & 0.0110 & 0.4938  \\
       3.7 & 38 (4480) &0.85\%& 0.0319 & 0.0056 & 0.0052 & 0.1110  \\
       4.2 & 33 (4480) &0.74\%& 0.0272 & 0.0061 & 0.0049 & 0.2997 \\
       5.0 & 45 (4480) &1.00\%& 0.0698 & 0.0552 & 0.0255 & 0.9022  \\
       6.0 & 30 (4480) &0.67\%& 0.0119 & -0.0028 & 0.0000 & 0.0981  \\
       $\overline{x}$ & 42.25 &0.94\% & 0.0341 & 0.012375 & 0.0092 & 0.3157 \\
       $\sigma$ & 9.02 &0.20\%& 0.0151 & 0.0169 & 0.0070 & 0.2617\\
       \bottomrule
    \end{tabular}}
    \begin{tablenotes} %添加此处
        \item The unit of excitation energy deviations are in eV, and bond length is in bohr. 
     \end{tablenotes}
\end{table}

%Fig.~\ref{fig7} (a) shows that the PES of the qEOM-SDt method almost coincides with that of the CASCI and qEOM-SDT methods despite incorporating an average of 16 (1\%) triple excitations across various bond strengths. Fig. \ref{fig7} (b) demonstrates the energy errors calculated by the three methods. A comparison between the curves obtained from the qEOM-SD and qEOM-SDt methods reveals a significant reduction in energy error with the qEOM-SDt method.

\subsection{2\,$^1Ag$ excited state of $\rm H_8$}\label{$H_8$: energy of 2 $^1Ag$ excited state}
\begin{figure}[H] 
    \centering
    \makebox[\textwidth][c]{\includegraphics[width=0.4\textwidth]{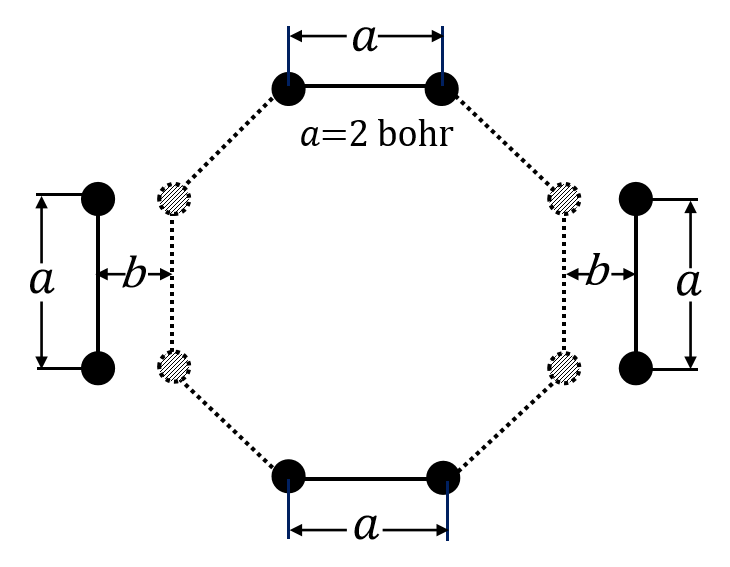}}
    \caption{\centering{Structure of the $\rm{H_8}$ system.
}}
    \label{fig6}
\end{figure}

It is well known that the EOM-CCSD method is able to treat the excited states dominated by single electron excitation while its accuracy rapidly deteriorates when double electron excitations are involved. The second $^1Ag$ state of the $\rm H_8$ system is dominated by the biexcited $2\,a_g^2 \rightarrow 1\,b_{1g}^2$ configuration, posing a great challenge for the EOM-CCSD and qEOM-SD methods. \cite{kowalski2001active} The structure of the $\rm H_8$ is depicted in Fig. \ref{fig6}, with its stretching distance from the regular octahedron denoted as $b$. 

The basis set for the hydrogen atom comprises three primitive Gaussian-type functions (S-type), with exponents of 4.50038, 0.681277, and 0.151374, and corresponding contraction coefficients of 0.07048, 0.40789, and 0.64767, respectively. \cite{jankowski1985davidson} The highest point symmetry $D_{2h}$ is used to reduce the number of operators in $\hat{\mathbf{R}}$. 

\begin{table}[!htbp]

\caption{\centering Energy deviations of the 2\,$^1Ag$ excited state computed with qEOM-SDt, qEOM-SD(t), qEOM-SDT and qEOM-SD methods for the $\rm H_8$ system at different $b$. The reference values are computed with the CASCI method.}
\label{table5}
\centering
\renewcommand{\arraystretch}{1.45}
\resizebox{\linewidth}{!}{
\begin{tabular}{ccccccc}
   \toprule
   \midrule
   $b$ & \textbf{Op} in $\hat{\mathbf{R}}_3$ after (before) & \%  & $\Delta E_{\rm qEOM-SDt}$ &$\Delta E_{\rm qEOM-SD(t)}$& $\Delta E_{\rm qEOM-SDT}$ & $\Delta E_{\rm qEOM-SD}$  \\
   % \midrule
   \Xhline{1pt}  %1pt代表粗细
   1.0 &  96 (1184) &$8.11\%$  & 0.1456 & 0.1326 & 0.0586 & 0.6989 \\
   
   0.8 &  88 (1184) &$7.43\%$& 0.1468 & 0.1347 & 0.0649 & 0.8357 \\

   0.6 &  90 (1184) &$7.60\%$  & 0.1465 & 0.1377 & 0.0813 & 0.9336 \\

   0.4 &  86 (1184) &$7.26\%$  & 0.1575 & 0.1493 & 0.1038 & 1.0000 \\

   0.2 &  84 (1184) &$7.09\%$  & 0.1757 & 0.1681 & 0.1310 &  0.9994 \\
   $\overline{x}$ & 88.80 & 7.50\% & 0.1544 & 0.1445 & 0.0879 & 0.8935\\
   $\sigma$ & 4.12 & 0.35\% & 0.0115 & 0.0131 & 0.0266 & 0.1143\\
   \bottomrule
\end{tabular}}
\begin{tablenotes} %添加此处
    \item The unit of excitation energy deviations are in eV, and $b$ is in bohr. 
 \end{tablenotes} %添加此处
\end{table}

Table~\ref{table5} shows the energy deviations of the 2\,$^1Ag$ excited state computed with qEOM-SD(t), qEOM-SDt, qEOM-SDT and qEOM-SD methods for the $\rm H_8$ system, which generally exhibit an increasing trend as $b$ decreases. The average energy deviation of qEOM-SD is about 0.89 eV for $b$ ranging from 1.0 to 0.2 bohr. However, for qEOM-SDt and qEOM-SD(t) methods, it is nearly 0.15 eV. In contrast with 2\,$^1\Sigma$ excited state of HF and 2\,$^1 \Delta$ excited state of $\rm CH^+$, more triple excitation operators up to an average of 7.50\% are involved in the qEOM-SDt and qEOM-SD(t) calculations. 
%that at each bond length, the energy calculated by the qEOM-SDt method achieves the same precision as that of the qEOM-SDT method, and is improved by nearly one order of magnitude compared to that of the qEOM-SD method. Moreover, by incorporating an average of 2.57\% of triple excitations in  the qEOM-SDt method at various stretching distances, we can cover more than 90\% of the energy correction contributed by all triple excitations.  

\section{Quantum Subspace Expansion with single, double, and triple excitations}\label{section5}
Both the QSE and qEOM methods have been suggested for calculating molecular excitation energies. These two methods also share some similar features in the subspace expansion and formulation of the working equation. Here, we extend the operator screening technique to the QSE to include the triple excitations.

\subsection{QSE-SDT method}
In the QSE method, the generalized eigenvalue equation is written as
\begin{equation}\label{eq11}
\boldsymbol{HD}_m=E_m\boldsymbol{SD}_m
\end{equation}
where $\boldsymbol{S}$ is the overlap matrix. The elements of $\boldsymbol{S}$ and $\boldsymbol{H}$ are defined as
\begin{equation}
\begin{split}
  \boldsymbol{S}_{IJ}&=\langle \psi_0|\hat{r}_I^{\dag} \hat{r}_J|\psi_{0}\rangle\\
  \boldsymbol{H}_{IJ}&=\langle \psi_0|\hat{r}_I^{\dag} \hat{H} \hat{r}_J|\psi_{0}\rangle
\end{split}
\end{equation}
%Equations \eqref{eq4} and \eqref{eq11}  represent the matrix forms of the Schrödinger equation under two different sets of bases. In the EOM method, the basis set $\{\hat{U}\hat{r}_I|{\psi_{\rm HF}}\rangle|\hat{r}_I \in \hat{R}\}$ ensures orthogonality between the ground state and excited states. 
In contrast to the qEOM method, the QSE method does not ensure the orthogonality among the subspace basis $\{\hat{r}_I\hat{U}|{\psi_{\rm HF}}\rangle|\hat{r}_I \in \hat{\mathbf{R}}\}$, so the ground state $|{\psi_{0}}\rangle$ should be included in the subspace expansion. 

As discussed in Section~\ref{section:Reducing operators by symmetry}, the working equation of QSE can also be block diagonalized, that is 
%\begin{equation}\nonumber
%{\rm Irrep}(\hat{T}_i|{\psi_{\rm HF}\rangle})=\rm Irrep(|\psi_{HF}\rangle)
%\end{equation}

%\noindent and
\begin{equation}\nonumber
{\rm Irrep}(\hat{r}_I|{\psi_{0}\rangle})=\rm Irrep(|\psi_m\rangle)
\end{equation}

Additionally, one can use the perturbation theory to further reduce the number of triple excitation operators in $\hat{\mathbf{R}}$ per irreducible representation. Like the qEOM, it is able to define the zero- and first-order wave function using $\hat{\mathbf{R}}_1 \cup \hat{\mathbf{R}}_2$ and  $\hat{\mathbf{R}}_3$ respectively as
\begin{equation}
\begin{split}
{|\psi_{m}^{(0)}\rangle}= \sum_{\substack{\hat{r}_I\in {\hat{\mathbf{R}}_{1}\cup\hat{\mathbf{R}}_{2}}}} \gamma_I\hat{r}_I|\psi_{0}\rangle \\
        {|\psi_{m}^{(1)}\rangle}= \sum_{\substack{\hat{r}_I\in {\hat{\mathbf{R}}_{3}}}} \gamma_I\hat{r}_I|\psi_{0}\rangle.
\end{split}
\end{equation}
As such, the weight coefficients can be  formulated following the derivation presented in Section~\ref{sec:screening} as
\begin{equation}\label{eq:W}
W_I=\frac{|\langle \psi_{\rm 0}|r_I^{\dagger}\hat{H}|\psi_{m}^{(0)}\rangle-E_m^{(0)} \langle \psi_{\rm 0}| r_I^{\dagger}|\psi_{m}^{(0)}\rangle|^{2}}{E_m^{(0)} \langle \psi_{\rm 0}|r_I^{\dagger} \hat{r}_I|\psi_{\rm 0}\rangle - \langle \psi_{\rm 0}|r_I^{\dagger}\hat{H}\hat{r}_I|\psi_{\rm 0}\rangle}
\end{equation}
Finally, the perturbation correction is performed as in Section~\ref{sec:screening} (4).

\subsection{2\,$^1Ag$ excited state of $\rm H_8$}\label{sec:QSE}
%The 2\,$^1Ag$ excited state of the $\rm H_8$ molecule cannot be accurately computed using the QSE method considering only the quadratic response subspace. We attempt to employ QSE+t for this calculation. The configuration of the $\rm H_8$ molecule is same as Section \ref{$H_8$: energy of 2 $^1Ag$ excited state}. The result of symmetry reduction process is the same as that of the EOM method. Table \ref{table6} displays the number of final remained triple operators and the energies error of the target state calculated by different method at various streching distances.
Table~\ref{table6} shows the energy deviations of the 2\,$^1Ag$ excited state computed with QSE-SDt, QSE-SD(t), QSE-SDT and QSE-SD methods for the $\rm H_8$ system. To facilitate comparison with the qEOM method, the basis set, molecular information, and the number of selected operators for each bond length are the same as those used in the qEOM calculations. The average energy deviations of QSE-SDt, QSE-SD(t), QSE-SDT and QSE-SD are about 0.15 eV, 0.14 eV, 0.09 eV and 0.89 eV. The overall performance of the QSE and qEOM methods is very similar. However, the QSE offers an improvement in accuracy by an order of magnitude of $10^{-3}$ compared to the qEOM method. 

\begin{table}[!htbp]
    \caption{\centering Energy deviations of the 2\,$^1Ag$ excited state computed with QSE-SDt, QSE-SD(t), QSE-SDT and QSE-SD methods for the $\rm H_8$ system at different $b$. The reference values are computed with the CASCI method.}
    \label{table6}
    \centering
    \renewcommand{\arraystretch}{1.45}
    \resizebox{\linewidth}{!}{
    \begin{tabular}{ccccccc}
       \toprule
       \midrule
       $b$ & \textbf{Op} in $\hat{\mathbf{R}}_3$ after (before)  &$\%$ & $\Delta E_{\rm QSE-SDt}$ &$\Delta E_{\rm QSE-SD(t)}$& $\Delta E_{\rm QSE-SDT}$ & $\Delta E_{\rm QSE-SD}$  \\
       % \midrule
       \Xhline{1pt}  %1pt代表粗细
   1.0 &  96 (1184) &$8.11\%$  & 0.1336 & 0.1235 & 0.0462& 0.6833 \\
   
   0.8 &  88 (1184) &$7.43\%$& 0.1446 &0.1341&0.0641  & 0.8331 \\

   0.6 &  90 (1184) &$7.60\%$  & 0.1464 & 0.1381 & 0.0816 & 0.9307\\

   0.4 &  86 (1184) &$7.26\%$  & 0.1560 & 0.1484 & 0.1023 & 0.9935 \\

   0.2 &  84 (1184) &$7.09\%$  & 0.1799 & 0.1732 & 0.1333 &  0.9899 \\
   $\overline{x}$ & 88.8 & $7.50\%$ & 0.1521 & 0.14346 & 0.0855 & 0.8861\\
   $\sigma$ & 4.1 & $0.35\%$ & 0.0156 & 0.0169 & 0.0303 & 0.1168\\
      
       \bottomrule
    \end{tabular}}
    \begin{tablenotes} %添加此处
    \item The units of energy and energy difference are eV and $b$ is in bohr. 
 \end{tablenotes} %添加此处
\end{table}

%At each bond length, by adding an average of $2.48\%$ of the triple excitations is sufficient to cover more than 90\% of the energy correction contributed by all triple excitations. Furthermore, the achieved energy accuracy matches that obtained by considering all triple excitations, surpassing the QSE results based solely on single and double excitations by an order of magnitude. The number of the remained operators and the computed energy accuracy of the QSE+t method are also comparable to those of the qEOM-SDt method.

\section{Conclusion and outlook} \label{sec:Conclusion}
In this work, we numerically demonstrate that the qEOM-SD method is unable to accurately describe molecule excitation energies dominated by double and high-order excitations, which commonly exist in the excited states of (near-) degenerate systems. In order to overcome this issue, we propose the qEOM-SDT method to improve the accuracy of the qEOM thoery by including the triple excitations. Numerical calculations of excitation energies for small molecular systems, including CH$^\dagger$, HF and H$_8$, demonstrates the qEOM-SDT method reduces the energe errors by one to two order of magnitude, in contrast to the qEOM-SD method.  

In order to reduce the computational complexity of qEOM-SDT, we propose a method called qEOM-SDt, which utilizes point group symmetry and perturbation theory to reduce the number of triple excitation operators in qEOM-SDT calculations. We further introduce qEOM-SD(t), where the energy corrections from the unselected triple excitation operators are replaced by their associated first-order perturbation energies. Numerical results show that qEOM-SD(t) generally provides better excitation state accuracy than qEOM-SDt. However, unlike qEOM-SDt and qEOM-SDT, it may underestimate excitation energies. Both methods can provide satisfactory excitation energies with few triple excitation operators, and in some cases, the excitation energy accuracy is higher than the ground state accuracy obtained from VQE calculations. 
%For $\rm{CH}^+$, HF, and $\rm{H}_8$ molecules, only an average of 1.98\%, 0.94\%, and 7.50\% of the total number of triple excitation operators are retained in the qEOM-SDt and qEOM-SD(t) methods. For these three molecules, the average energy errors of qEOM-SDt are 0.20, 0.03, and 0.15 eV, respectively, while those of qEOM-SDT are 0.17, 0.012, and 0.14 eV. So, 
The computational bottleneck of the qEOM-SDT mainly results from estimating the weight coefficients used for the operator screening instead of measuring all Hamiltonian matrix elements involving triple excitation operators in the qEOM-SDT. 
The same operator reduction procedure is also extended to the QSE for calculating excitation energies. The new algorithms presented in this work provide a promising tool for studying electronically excited states on a quantum computer.

\section{Acknowledgments}
This work was supported by Innovation Program for Quantum Science and Technology (2021ZD0303306), the National Natural Science Foundation of China (22073086, 22288201), Anhui Initiative in Quantum Information Technologies (AHY090400), and the Fundamental Research Funds for the Central Universities (WK2060000018).

\section{Appendix}
In the following, we will show the derivation of the criterion of operator selection in ADAPT-VQE ansatz with UCC operator pool. This mainly follows the derivation of the criterion of operator selection in ADAPT-VQE ansatz showed in \cite{cao2022progress}.

The point group $G$ of the system is Abelian, and $\hat{R}_i$ is an symmetry operation in it. So, we have $\hat{R}_i^{\dag}\hat{H}\hat{R}_i=\hat{H}$, and
 \begin{gather}
    \hat{H}\hat{R}_i|\psi_0\rangle = \hat{R}_i\hat{H}|\psi_0\rangle = E_0\hat{R}_i|\psi_0\rangle
\end{gather}
If the ground state is nondegenerate, $\hat{R}_i|\psi_0\rangle$ and  $|\psi_0\rangle$ is the same state but can differ by a coefficient, i.e.,
\begin{gather}
\hat{R}_i|\psi_0\rangle = \lambda|\psi_0\rangle
\end{gather}
Thus, $|\psi_0\rangle$ forms an one-dimensional invariant subspace under the point group $G$, and in this subspace the irreducible character correspond to $\hat{R}_i$ is $\lambda$. The HF state also forms an invariant subspace, and its irreducible character for $\hat{R}_i$ is termed as $\beta$.
 \begin{gather}
\hat{R}_i|\psi_{\rm{HF}}\rangle = \beta|\psi_{\rm{HF}}\rangle
\end{gather}
Since $\hat{R}_i^{\dag}\hat{R}_i = \hat I$, we have
\begin{gather}
\langle \psi_{\rm{HF}}|\psi_0\rangle = \langle \psi_{\rm{HF}}|\hat{R}_i^{\dag}\hat{R}_i|\psi_0\rangle = \lambda\beta\langle \psi_{\rm{HF}}|\psi_0\rangle
\end{gather}
So, $\lambda\beta = 1$. And for the Abelian group, the characters are 1 or -1. Therefore, $\lambda = \beta$, which means $\psi_{\rm{HF}}$ and $\psi_0$ belong to the same irreducible representation (irrep).
\begin{gather}\label{irrep_same}
\rm{Irrep}(\psi_{\rm{HF}}) = \rm{Irrep}(\psi_0)
\end{gather}

The ground state calculated by ADAPT-VQE ansatz
 \begin{gather}
    |\psi_{\rm{VQE}}\rangle = \prod \limits_{i=0}
     \left(e^{\theta_i\left(\hat{T}_i-\hat{T}_i^{\dag}\right)}\right)|\psi_{\rm HF}\rangle
\end{gather}
should satisfy the above condition. Based on Taylor expansion and $\hat{T}_i^{\dag}\psi_{\rm{HF}} = 0$, we have
 \begin{gather}
 \begin{split}
    |\psi_{\rm{VQE}}\rangle &= \prod \limits_{i=0}
     \left(1+\theta_i(\hat{T}_i-\hat{T}_i^{\dag})+\frac{\theta_i^2}{2}(\hat{T}_i-\hat{T}_i^{\dag})^2+\cdots\right)|\psi_{\rm HF}\rangle\\
     &= c_0|\psi_{\rm{HF}}\rangle + \sum_{i}c_iT_i|\psi_{\rm{HF}}\rangle +\sum_{i}c_iT_i^2|\psi_{\rm {HF}}\rangle+\cdots
\end{split}
\end{gather}
Since $\psi_{\rm{HF}}$ and $\psi_0$ belong to the same irrep, every term in the above equation should belong to the same irrep of $\psi_{\rm{HF}}$, i.e.,
\begin{gather}
{\rm{Irrep}} (T_i\psi_{\rm{HF}}) = \rm{Irrep}(\psi_0)
\end{gather}

\footnotesize{
\bibliography{qc}

\providecommand{\latin}[1]{#1}
\makeatletter
\providecommand{\doi}
  {\begingroup\let\do\@makeother\dospecials
  \catcode`\{=1 \catcode`\}=2 \doi@aux}
\providecommand{\doi@aux}[1]{\endgroup\texttt{#1}}
\makeatother
\providecommand*\mcitethebibliography{\thebibliography}
\csname @ifundefined\endcsname{endmcitethebibliography}  {\let\endmcitethebibliography\endthebibliography}{}
\begin{mcitethebibliography}{40}
\providecommand*\natexlab[1]{#1}
\providecommand*\mciteSetBstSublistMode[1]{}
\providecommand*\mciteSetBstMaxWidthForm[2]{}
\providecommand*\mciteBstWouldAddEndPuncttrue
  {\def\EndOfBibitem{\unskip.}}
\providecommand*\mciteBstWouldAddEndPunctfalse
  {\let\EndOfBibitem\relax}
\providecommand*\mciteSetBstMidEndSepPunct[3]{}
\providecommand*\mciteSetBstSublistLabelBeginEnd[3]{}
\providecommand*\EndOfBibitem{}
\mciteSetBstSublistMode{f}
\mciteSetBstMaxWidthForm{subitem}{(\alph{mcitesubitemcount})}
\mciteSetBstSublistLabelBeginEnd
  {\mcitemaxwidthsubitemform\space}
  {\relax}
  {\relax}

\bibitem[Pulay(2011)]{Pul11}
Pulay,~P. A perspective on the CASPT2 method. \emph{Int.\ J. Quantum Chem.} \textbf{2011}, \emph{111}, 3273--3279\relax
\mciteBstWouldAddEndPuncttrue
\mciteSetBstMidEndSepPunct{\mcitedefaultmidpunct}
{\mcitedefaultendpunct}{\mcitedefaultseppunct}\relax
\EndOfBibitem
\bibitem[Szalay \latin{et~al.}(2012)Szalay, M\"uller, Gidofalvi, Lischka, and Shepard]{SzaMulGid12}
Szalay,~P.~G.; M\"uller,~T.; Gidofalvi,~G.; Lischka,~H.; Shepard,~R. Multiconfiguration Self-Consistent Field and Multireference Configuration Interaction Methods and Applications. \emph{Chem.\ Rev.} \textbf{2012}, \emph{112}, 108--181\relax
\mciteBstWouldAddEndPuncttrue
\mciteSetBstMidEndSepPunct{\mcitedefaultmidpunct}
{\mcitedefaultendpunct}{\mcitedefaultseppunct}\relax
\EndOfBibitem
\bibitem[Baiardi \latin{et~al.}(2022)Baiardi, Kelemen, and Reiher]{BaiKelRei22}
Baiardi,~A.; Kelemen,~A.~K.; Reiher,~M. Excited-State DMRG Made Simple with FEAST. \emph{J.~Chem.\ Theory Comput.} \textbf{2022}, \emph{18}, 415--430\relax
\mciteBstWouldAddEndPuncttrue
\mciteSetBstMidEndSepPunct{\mcitedefaultmidpunct}
{\mcitedefaultendpunct}{\mcitedefaultseppunct}\relax
\EndOfBibitem
\bibitem[Schriber and Evangelista(2017)Schriber, and Evangelista]{SchEva17}
Schriber,~J.~B.; Evangelista,~F.~A. Adaptive Configuration Interaction for Computing Challenging Electronic Excited States with Tunable Accuracy. \emph{J.~Chem.\ Theory Comput.} \textbf{2017}, \emph{13}, 5354--5366\relax
\mciteBstWouldAddEndPuncttrue
\mciteSetBstMidEndSepPunct{\mcitedefaultmidpunct}
{\mcitedefaultendpunct}{\mcitedefaultseppunct}\relax
\EndOfBibitem
\bibitem[Aspuru-Guzik \latin{et~al.}(2005)Aspuru-Guzik, Dutoi, Love, and Head-Gordon]{aspuru2005simulated}
Aspuru-Guzik,~A.; Dutoi,~A.~D.; Love,~P.~J.; Head-Gordon,~M. Simulated quantum computation of molecular energies. \emph{Science} \textbf{2005}, \emph{309}, 1704--1707\relax
\mciteBstWouldAddEndPuncttrue
\mciteSetBstMidEndSepPunct{\mcitedefaultmidpunct}
{\mcitedefaultendpunct}{\mcitedefaultseppunct}\relax
\EndOfBibitem
\bibitem[Peruzzo \latin{et~al.}(2014)Peruzzo, McClean, Shadbolt, Yung, Zhou, Love, Aspuru-Guzik, and O’brien]{peruzzo2014variational}
Peruzzo,~A.; McClean,~J.; Shadbolt,~P.; Yung,~M.-H.; Zhou,~X.-Q.; Love,~P.~J.; Aspuru-Guzik,~A.; O’brien,~J.~L. A variational eigenvalue solver on a photonic quantum processor. \emph{Nature communications} \textbf{2014}, \emph{5}, 4213\relax
\mciteBstWouldAddEndPuncttrue
\mciteSetBstMidEndSepPunct{\mcitedefaultmidpunct}
{\mcitedefaultendpunct}{\mcitedefaultseppunct}\relax
\EndOfBibitem
\bibitem[McArdle \latin{et~al.}(2020)McArdle, Endo, Aspuru-Guzik, Benjamin, and Yuan]{mcardle2020quantum}
McArdle,~S.; Endo,~S.; Aspuru-Guzik,~A.; Benjamin,~S.~C.; Yuan,~X. Quantum computational chemistry. \emph{Reviews of Modern Physics} \textbf{2020}, \emph{92}, 015003\relax
\mciteBstWouldAddEndPuncttrue
\mciteSetBstMidEndSepPunct{\mcitedefaultmidpunct}
{\mcitedefaultendpunct}{\mcitedefaultseppunct}\relax
\EndOfBibitem
\bibitem[Su \latin{et~al.}(2021)Su, Berry, Wiebe, Rubin, and Babbush]{su2021fault}
Su,~Y.; Berry,~D.~W.; Wiebe,~N.; Rubin,~N.; Babbush,~R. Fault-tolerant quantum simulations of chemistry in first quantization. \emph{PRX Quantum} \textbf{2021}, \emph{2}, 040332\relax
\mciteBstWouldAddEndPuncttrue
\mciteSetBstMidEndSepPunct{\mcitedefaultmidpunct}
{\mcitedefaultendpunct}{\mcitedefaultseppunct}\relax
\EndOfBibitem
\bibitem[O’Malley \latin{et~al.}(2016)O’Malley, Babbush, Kivlichan, Romero, McClean, Barends, Kelly, Roushan, Tranter, Ding, \latin{et~al.} others]{o2016scalable}
O’Malley,~P.~J.; Babbush,~R.; Kivlichan,~I.~D.; Romero,~J.; McClean,~J.~R.; Barends,~R.; Kelly,~J.; Roushan,~P.; Tranter,~A.; Ding,~N.; others Scalable quantum simulation of molecular energies. \emph{Physical Review X} \textbf{2016}, \emph{6}, 031007\relax
\mciteBstWouldAddEndPuncttrue
\mciteSetBstMidEndSepPunct{\mcitedefaultmidpunct}
{\mcitedefaultendpunct}{\mcitedefaultseppunct}\relax
\EndOfBibitem
\bibitem[Parrish \latin{et~al.}(2019)Parrish, Hohenstein, McMahon, and Mart{\'\i}nez]{parrish2019quantum}
Parrish,~R.~M.; Hohenstein,~E.~G.; McMahon,~P.~L.; Mart{\'\i}nez,~T.~J. Quantum computation of electronic transitions using a variational quantum eigensolver. \emph{Physical review letters} \textbf{2019}, \emph{122}, 230401\relax
\mciteBstWouldAddEndPuncttrue
\mciteSetBstMidEndSepPunct{\mcitedefaultmidpunct}
{\mcitedefaultendpunct}{\mcitedefaultseppunct}\relax
\EndOfBibitem
\bibitem[Nakanishi \latin{et~al.}(2019)Nakanishi, Mitarai, and Fujii]{nakanishi2019subspace}
Nakanishi,~K.~M.; Mitarai,~K.; Fujii,~K. Subspace-search variational quantum eigensolver for excited states. \emph{Physical Review Research} \textbf{2019}, \emph{1}, 033062\relax
\mciteBstWouldAddEndPuncttrue
\mciteSetBstMidEndSepPunct{\mcitedefaultmidpunct}
{\mcitedefaultendpunct}{\mcitedefaultseppunct}\relax
\EndOfBibitem
\bibitem[Ryabinkin \latin{et~al.}(2018)Ryabinkin, Genin, and Izmaylov]{ryabinkin2018constrained}
Ryabinkin,~I.~G.; Genin,~S.~N.; Izmaylov,~A.~F. Constrained variational quantum eigensolver: Quantum computer search engine in the Fock space. \emph{Journal of chemical theory and computation} \textbf{2018}, \emph{15}, 249--255\relax
\mciteBstWouldAddEndPuncttrue
\mciteSetBstMidEndSepPunct{\mcitedefaultmidpunct}
{\mcitedefaultendpunct}{\mcitedefaultseppunct}\relax
\EndOfBibitem
\bibitem[Stair \latin{et~al.}(2020)Stair, Huang, and Evangelista]{stair2020multireference}
Stair,~N.~H.; Huang,~R.; Evangelista,~F.~A. A multireference quantum Krylov algorithm for strongly correlated electrons. \emph{J.~Chem.\ Theory Comput.} \textbf{2020}, \emph{16}, 2236--2245\relax
\mciteBstWouldAddEndPuncttrue
\mciteSetBstMidEndSepPunct{\mcitedefaultmidpunct}
{\mcitedefaultendpunct}{\mcitedefaultseppunct}\relax
\EndOfBibitem
\bibitem[Jones \latin{et~al.}(2019)Jones, Endo, McArdle, Yuan, and Benjamin]{jones2019variational}
Jones,~T.; Endo,~S.; McArdle,~S.; Yuan,~X.; Benjamin,~S.~C. Variational quantum algorithms for discovering Hamiltonian spectra. \emph{Physical Review A} \textbf{2019}, \emph{99}, 062304\relax
\mciteBstWouldAddEndPuncttrue
\mciteSetBstMidEndSepPunct{\mcitedefaultmidpunct}
{\mcitedefaultendpunct}{\mcitedefaultseppunct}\relax
\EndOfBibitem
\bibitem[Tilly \latin{et~al.}(2020)Tilly, Jones, Chen, Wossnig, and Grant]{tilly2020computation}
Tilly,~J.; Jones,~G.; Chen,~H.; Wossnig,~L.; Grant,~E. Computation of molecular excited states on IBM quantum computers using a discriminative variational quantum eigensolver. \emph{Physical Review A} \textbf{2020}, \emph{102}, 062425\relax
\mciteBstWouldAddEndPuncttrue
\mciteSetBstMidEndSepPunct{\mcitedefaultmidpunct}
{\mcitedefaultendpunct}{\mcitedefaultseppunct}\relax
\EndOfBibitem
\bibitem[Bauman \latin{et~al.}(2021)Bauman, Liu, Bylaska, Krishnamoorthy, Low, Granade, Wiebe, Baker, Peng, Roetteler, Troyer, and Kowalski]{BauLiuByl21}
Bauman,~N.~P.; Liu,~H.; Bylaska,~E.~J.; Krishnamoorthy,~S.; Low,~G.~H.; Granade,~C.~E.; Wiebe,~N.; Baker,~N.~A.; Peng,~B.; Roetteler,~M.; Troyer,~M.; Kowalski,~K. Toward Quantum Computing for High-Energy Excited States in Molecular Systems: Quantum Phase Estimations of Core-Level States. \emph{J.~Chem.\ Theory Comput.} \textbf{2021}, \emph{17}, 201--210\relax
\mciteBstWouldAddEndPuncttrue
\mciteSetBstMidEndSepPunct{\mcitedefaultmidpunct}
{\mcitedefaultendpunct}{\mcitedefaultseppunct}\relax
\EndOfBibitem
\bibitem[Kang \latin{et~al.}(2022)Kang, Bauman, Krishnamoorthy, and Kowalski]{KanBauKri22}
Kang,~C.; Bauman,~N.~P.; Krishnamoorthy,~S.; Kowalski,~K. Optimized Quantum Phase Estimation for Simulating Electronic States in Various Energy Regimes. \emph{J.~Chem.\ Theory Comput.} \textbf{2022}, \emph{18}, 6567--6576\relax
\mciteBstWouldAddEndPuncttrue
\mciteSetBstMidEndSepPunct{\mcitedefaultmidpunct}
{\mcitedefaultendpunct}{\mcitedefaultseppunct}\relax
\EndOfBibitem
\bibitem[Baker(2021)]{baker2021lanczos}
Baker,~T.~E. Lanczos recursion on a quantum computer for the Green's function and ground state. \emph{Physical Review A} \textbf{2021}, \emph{103}, 032404\relax
\mciteBstWouldAddEndPuncttrue
\mciteSetBstMidEndSepPunct{\mcitedefaultmidpunct}
{\mcitedefaultendpunct}{\mcitedefaultseppunct}\relax
\EndOfBibitem
\bibitem[Preskill(2018)]{preskill2018quantum}
Preskill,~J. Quantum computing in the NISQ era and beyond. \emph{Quantum} \textbf{2018}, \emph{2}, 79\relax
\mciteBstWouldAddEndPuncttrue
\mciteSetBstMidEndSepPunct{\mcitedefaultmidpunct}
{\mcitedefaultendpunct}{\mcitedefaultseppunct}\relax
\EndOfBibitem
\bibitem[Sharma \latin{et~al.}(2020)Sharma, Khatri, Cerezo, and Coles]{2020Noise}
Sharma,~K.; Khatri,~S.; Cerezo,~M.; Coles,~P.~J. Noise resilience of variational quantum compiling. \emph{New Journal of Physics} \textbf{2020}, \emph{22}, 043006 (29pp)\relax
\mciteBstWouldAddEndPuncttrue
\mciteSetBstMidEndSepPunct{\mcitedefaultmidpunct}
{\mcitedefaultendpunct}{\mcitedefaultseppunct}\relax
\EndOfBibitem
\bibitem[Chan \latin{et~al.}(2021)Chan, Fitzpatrick, Segarra-Mart{\'\i}, Bearpark, and Tew]{chan2021molecular}
Chan,~H. H.~S.; Fitzpatrick,~N.; Segarra-Mart{\'\i},~J.; Bearpark,~M.~J.; Tew,~D.~P. Molecular excited state calculations with adaptive wavefunctions on a quantum eigensolver emulation: reducing circuit depth and separating spin states. \emph{Physical Chemistry Chemical Physics} \textbf{2021}, \emph{23}, 26438--26450\relax
\mciteBstWouldAddEndPuncttrue
\mciteSetBstMidEndSepPunct{\mcitedefaultmidpunct}
{\mcitedefaultendpunct}{\mcitedefaultseppunct}\relax
\EndOfBibitem
\bibitem[Higgott \latin{et~al.}(2019)Higgott, Wang, and Brierley]{higgott2019variational}
Higgott,~O.; Wang,~D.; Brierley,~S. Variational quantum computation of excited states. \emph{Quantum} \textbf{2019}, \emph{3}, 156\relax
\mciteBstWouldAddEndPuncttrue
\mciteSetBstMidEndSepPunct{\mcitedefaultmidpunct}
{\mcitedefaultendpunct}{\mcitedefaultseppunct}\relax
\EndOfBibitem
\bibitem[Santagati \latin{et~al.}(2018)Santagati, Wang, Gentile, Paesani, Wiebe, McClean, Morley-Short, Shadbolt, Bonneau, Silverstone, \latin{et~al.} others]{santagati2018witnessing}
Santagati,~R.; Wang,~J.; Gentile,~A.~A.; Paesani,~S.; Wiebe,~N.; McClean,~J.~R.; Morley-Short,~S.; Shadbolt,~P.~J.; Bonneau,~D.; Silverstone,~J.~W.; others Witnessing eigenstates for quantum simulation of Hamiltonian spectra. \emph{Science advances} \textbf{2018}, \emph{4}, eaap9646\relax
\mciteBstWouldAddEndPuncttrue
\mciteSetBstMidEndSepPunct{\mcitedefaultmidpunct}
{\mcitedefaultendpunct}{\mcitedefaultseppunct}\relax
\EndOfBibitem
\bibitem[Lee \latin{et~al.}(2018)Lee, Huggins, Head-Gordon, and Whaley]{lee2018generalized}
Lee,~J.; Huggins,~W.~J.; Head-Gordon,~M.; Whaley,~K.~B. Generalized unitary coupled cluster wave functions for quantum computation. \emph{J.~Chem.\ Theory Comput.} \textbf{2018}, \emph{15}, 311--324\relax
\mciteBstWouldAddEndPuncttrue
\mciteSetBstMidEndSepPunct{\mcitedefaultmidpunct}
{\mcitedefaultendpunct}{\mcitedefaultseppunct}\relax
\EndOfBibitem
\bibitem[McClean \latin{et~al.}(2017)McClean, Kimchi-Schwartz, Carter, and De~Jong]{mcclean2017hybrid}
McClean,~J.~R.; Kimchi-Schwartz,~M.~E.; Carter,~J.; De~Jong,~W.~A. Hybrid quantum-classical hierarchy for mitigation of decoherence and determination of excited states. \emph{Physical Review A} \textbf{2017}, \emph{95}, 042308\relax
\mciteBstWouldAddEndPuncttrue
\mciteSetBstMidEndSepPunct{\mcitedefaultmidpunct}
{\mcitedefaultendpunct}{\mcitedefaultseppunct}\relax
\EndOfBibitem
\bibitem[Ollitrault \latin{et~al.}(2020)Ollitrault, Kandala, Chen, Barkoutsos, Mezzacapo, Pistoia, Sheldon, Woerner, Gambetta, and Tavernelli]{ollitrault2020quantum}
Ollitrault,~P.~J.; Kandala,~A.; Chen,~C.-F.; Barkoutsos,~P.~K.; Mezzacapo,~A.; Pistoia,~M.; Sheldon,~S.; Woerner,~S.; Gambetta,~J.~M.; Tavernelli,~I. Quantum equation of motion for computing molecular excitation energies on a noisy quantum processor. \emph{Physical Review Research} \textbf{2020}, \emph{2}, 043140\relax
\mciteBstWouldAddEndPuncttrue
\mciteSetBstMidEndSepPunct{\mcitedefaultmidpunct}
{\mcitedefaultendpunct}{\mcitedefaultseppunct}\relax
\EndOfBibitem
\bibitem[Asthana \latin{et~al.}(2023)Asthana, Kumar, Abraham, Grimsley, Zhang, Cincio, Tretiak, Dub, Economou, Barnes, \latin{et~al.} others]{asthana2023quantum}
Asthana,~A.; Kumar,~A.; Abraham,~V.; Grimsley,~H.; Zhang,~Y.; Cincio,~L.; Tretiak,~S.; Dub,~P.~A.; Economou,~S.~E.; Barnes,~E.; others Quantum self-consistent equation-of-motion method for computing molecular excitation energies, ionization potentials, and electron affinities on a quantum computer. \emph{Chemical Science} \textbf{2023}, \emph{14}, 2405--2418\relax
\mciteBstWouldAddEndPuncttrue
\mciteSetBstMidEndSepPunct{\mcitedefaultmidpunct}
{\mcitedefaultendpunct}{\mcitedefaultseppunct}\relax
\EndOfBibitem
\bibitem[Piecuch \latin{et~al.}(2004)Piecuch, Kowalski, Pimienta, Fan, Lodriguito, McGuire, Kucharski, Ku{\'s}, and Musia{\l}]{piecuch2004method}
Piecuch,~P.; Kowalski,~K.; Pimienta,~I.; Fan,~P.-D.; Lodriguito,~M.; McGuire,~M.; Kucharski,~S.; Ku{\'s},~T.; Musia{\l},~M. Method of moments of coupled-cluster equations: a new formalism for designing accurate electronic structure methods for ground and excited states. \emph{Theoretical Chemistry Accounts} \textbf{2004}, \emph{112}, 349--393\relax
\mciteBstWouldAddEndPuncttrue
\mciteSetBstMidEndSepPunct{\mcitedefaultmidpunct}
{\mcitedefaultendpunct}{\mcitedefaultseppunct}\relax
\EndOfBibitem
\bibitem[W{\l}och \latin{et~al.}(2005)W{\l}och, Gour, Kowalski, and Piecuch]{wloch2005extension}
W{\l}och,~M.; Gour,~J.~R.; Kowalski,~K.; Piecuch,~P. Extension of renormalized coupled-cluster methods including triple excitations to excited electronic states of open-shell molecules. \emph{The Journal of chemical physics} \textbf{2005}, \emph{122}\relax
\mciteBstWouldAddEndPuncttrue
\mciteSetBstMidEndSepPunct{\mcitedefaultmidpunct}
{\mcitedefaultendpunct}{\mcitedefaultseppunct}\relax
\EndOfBibitem
\bibitem[Grimsley \latin{et~al.}(2019)Grimsley, Economou, Barnes, and Mayhall]{grimsley2019adaptive}
Grimsley,~H.~R.; Economou,~S.~E.; Barnes,~E.; Mayhall,~N.~J. An adaptive variational algorithm for exact molecular simulations on a quantum computer. \emph{Nature communications} \textbf{2019}, \emph{10}, 3007\relax
\mciteBstWouldAddEndPuncttrue
\mciteSetBstMidEndSepPunct{\mcitedefaultmidpunct}
{\mcitedefaultendpunct}{\mcitedefaultseppunct}\relax
\EndOfBibitem
\bibitem[Cao \latin{et~al.}(2022)Cao, Hu, Zhang, Xu, Chen, Yu, Li, Hu, Lv, and Yung]{cao2022progress}
Cao,~C.; Hu,~J.; Zhang,~W.; Xu,~X.; Chen,~D.; Yu,~F.; Li,~J.; Hu,~H.-S.; Lv,~D.; Yung,~M.-H. Progress toward larger molecular simulation on a quantum computer: Simulating a system with up to 28 qubits accelerated by point-group symmetry. \emph{Physical Review A} \textbf{2022}, \emph{105}, 062452\relax
\mciteBstWouldAddEndPuncttrue
\mciteSetBstMidEndSepPunct{\mcitedefaultmidpunct}
{\mcitedefaultendpunct}{\mcitedefaultseppunct}\relax
\EndOfBibitem
\bibitem[Cotton(1991)]{cotton1991chemical}
Cotton,~F.~A. \emph{Chemical applications of group theory}; John Wiley \& Sons, 1991\relax
\mciteBstWouldAddEndPuncttrue
\mciteSetBstMidEndSepPunct{\mcitedefaultmidpunct}
{\mcitedefaultendpunct}{\mcitedefaultseppunct}\relax
\EndOfBibitem
\bibitem[Fan \latin{et~al.}(2022)Fan, Liu, Zeng, Xu, Shang, Li, and Yang]{JUSTC-2022-0118}
Fan,~Y.; Liu,~J.; Zeng,~X.; Xu,~Z.; Shang,~H.; Li,~Z.; Yang,~J. Q<sup>2</sup>Chemistry: A quantum computation platform for quantum chemistry. \emph{JUSTC} \textbf{2022}, \emph{52}, 2\relax
\mciteBstWouldAddEndPuncttrue
\mciteSetBstMidEndSepPunct{\mcitedefaultmidpunct}
{\mcitedefaultendpunct}{\mcitedefaultseppunct}\relax
\EndOfBibitem
\bibitem[Sun \latin{et~al.}(2018)Sun, Berkelbach, Blunt, Booth, Guo, Li, Liu, McClain, Sayfutyarova, Sharma, \latin{et~al.} others]{sun2018pyscf}
Sun,~Q.; Berkelbach,~T.~C.; Blunt,~N.~S.; Booth,~G.~H.; Guo,~S.; Li,~Z.; Liu,~J.; McClain,~J.~D.; Sayfutyarova,~E.~R.; Sharma,~S.; others PySCF: the Python-based simulations of chemistry framework. \emph{Wiley Interdisciplinary Reviews: Computational Molecular Science} \textbf{2018}, \emph{8}, e1340\relax
\mciteBstWouldAddEndPuncttrue
\mciteSetBstMidEndSepPunct{\mcitedefaultmidpunct}
{\mcitedefaultendpunct}{\mcitedefaultseppunct}\relax
\EndOfBibitem
\bibitem[McClean \latin{et~al.}(2020)McClean, Rubin, Sung, Kivlichan, Bonet-Monroig, Cao, Dai, Fried, Gidney, Gimby, \latin{et~al.} others]{mcclean2020openfermion}
McClean,~J.~R.; Rubin,~N.~C.; Sung,~K.~J.; Kivlichan,~I.~D.; Bonet-Monroig,~X.; Cao,~Y.; Dai,~C.; Fried,~E.~S.; Gidney,~C.; Gimby,~B.; others OpenFermion: the electronic structure package for quantum computers. \emph{Quantum Science and Technology} \textbf{2020}, \emph{5}, 034014\relax
\mciteBstWouldAddEndPuncttrue
\mciteSetBstMidEndSepPunct{\mcitedefaultmidpunct}
{\mcitedefaultendpunct}{\mcitedefaultseppunct}\relax
\EndOfBibitem
\bibitem[Virtanen \latin{et~al.}(2020)Virtanen, Gommers, Oliphant, Haberland, Reddy, Cournapeau, Burovski, Peterson, Weckesser, Bright, \latin{et~al.} others]{virtanen2020scipy}
Virtanen,~P.; Gommers,~R.; Oliphant,~T.~E.; Haberland,~M.; Reddy,~T.; Cournapeau,~D.; Burovski,~E.; Peterson,~P.; Weckesser,~W.; Bright,~J.; others SciPy 1.0: fundamental algorithms for scientific computing in Python. \emph{Nature methods} \textbf{2020}, \emph{17}, 261--272\relax
\mciteBstWouldAddEndPuncttrue
\mciteSetBstMidEndSepPunct{\mcitedefaultmidpunct}
{\mcitedefaultendpunct}{\mcitedefaultseppunct}\relax
\EndOfBibitem
\bibitem[Kowalski and Piecuch(2004)Kowalski, and Piecuch]{kowalski2004new}
Kowalski,~K.; Piecuch,~P. New coupled-cluster methods with singles, doubles, and noniterative triples for high accuracy calculations of excited electronic states. \emph{J.~Chem.\ Phys.} \textbf{2004}, \emph{120}, 1715--1738\relax
\mciteBstWouldAddEndPuncttrue
\mciteSetBstMidEndSepPunct{\mcitedefaultmidpunct}
{\mcitedefaultendpunct}{\mcitedefaultseppunct}\relax
\EndOfBibitem
\bibitem[Kowalski and Piecuch(2001)Kowalski, and Piecuch]{kowalski2001active}
Kowalski,~K.; Piecuch,~P. The active-space equation-of-motion coupled-cluster methods for excited electronic states: Full EOMCCSDt. \emph{The Journal of Chemical Physics} \textbf{2001}, \emph{115}, 643--651\relax
\mciteBstWouldAddEndPuncttrue
\mciteSetBstMidEndSepPunct{\mcitedefaultmidpunct}
{\mcitedefaultendpunct}{\mcitedefaultseppunct}\relax
\EndOfBibitem
\bibitem[Jankowski \latin{et~al.}(1985)Jankowski, Meissner, and Wasilewski]{jankowski1985davidson}
Jankowski,~K.; Meissner,~L.; Wasilewski,~J. Davidson-type corrections for quasidegenerate states. \emph{International journal of quantum chemistry} \textbf{1985}, \emph{28}, 931--942\relax
\mciteBstWouldAddEndPuncttrue
\mciteSetBstMidEndSepPunct{\mcitedefaultmidpunct}
{\mcitedefaultendpunct}{\mcitedefaultseppunct}\relax
\EndOfBibitem
\end{mcitethebibliography}
}

\vspace{3ex}

%\[
%\includegraphics[width=0.5\textwidth]{TOC.pdf}
%\]
%\centerline{TOC graphic}

\end{document}